# Valuation Models Applied to Value-Based Management—Application to the Case of UK Companies with Problems

**Marcel Ausloos** [1,2,3,*,†]

[1] School of Business, Brookfield Campus, University of Leicester Leicester, LE2 1RQ, Leicester, UK
[2] Department of Statistics and Econometrics, Bucharest University of Economic Studies, 6 Piata Romana, 1st district, 010374 Bucharest, Romania
[3] Physics, GRAPES, rue de la Belle Jardinière 483, B-4031 Liège, Fédération Wallonie-Bruxelles, Belgium
* Correspondence: ma683@le.ac.uk; marcel.ausloos@ase.ro;
 marcel.ausloos@uliege.be; marcel.ausloos@ulg.ac.be

**Abstract:** Many still rightly wonder whether accounting numbers affect business value. Basic questions are "why?" and "how? We aim at promoting an objective choice on how optimizing the most suitable valuation methods under a "value-based management" framework through some performance measurement systems. First, we present a comprehensive review of valuation methods. Three valuations methods, (i) Free Cash Flow Valuation Model (FCFVM), (ii) Residual Earning Valuation Model (REVM) and (iii) Abnormal Earning Growth Model (AEGM), are presented. We point out advantages and limitations. As applications, the proofs of our findings are illustrated on three study cases: Marks & Spencer's (M&S's) business pattern (size and growth prospect), which had a recently advertised valuation "problem", and two comparable companies, Tesco and Sainsbury's, all three chosen for multiple-based valuation. For the purpose, two value drivers are chosen, EnV/EBIT (entity value/earnings before interests and taxes) and the corresponding EnV/Sales. Thus, the question whether accounting numbers through models based on mathematical economics truly affect business value has an answer: "Maybe, yes". JEL: C52 Model Evaluation, Validation, and Selection; C02 Mathematical Methods; C18 Methodological Issues: General; D46 Value Theory; G32 Value of Firms; M41 Accounting.

## 1. Introduction

### 1.1. Value-Based Management—The Importance of Value Creation

Value-based management (VBM) represents one of the four evolutionary stages in econometrics accounting. Nowadays, the valuation of business not only provides the result of how much a corporation is worth on the market, but also proposes elements that stimulate the business' value itself (Martin et al. 2009). Consequently, a comprehensive framework known as "Value-Based Management" has emerged, which focuses on: (1) measuring company value; (2) designing and implementing strategies that provide the largest potential for shareholder value creation; (3) applying information systems that concentrate on value creation and the "driver" of value, throughout a corporation's business departments, product, and customer segments; (4) integrating business plan and resource allocation along with value creation; (5) approving the original performance measurement systems which reflect value creation (Ittner and Larcker 2001).

Our paper specifically focuses with such aspects of VBM, dealing with the econometrics performance measurement system in/of a corporation. Of course, different models can be applied in view of implementing VBM, such as the free cash flow method, the economic value added/market value added (EVA/MVA) method, and the cash flow return on investment approach. Among these methods, EVA is widely believed to be the most effective (Chen and Dodd 1997; Martin and Petty 2001; Young and O'Byrne 2000). For completeness, let it be mentioned that when comparing valuation efficiency between stock markets, in particular, the Mainland China and Hong Kong, Yi et al. (2019) find out that a data envelopment analysis (DEA) model is more suitable for measuring the relative valuation level and efficiency than the P/E ratio.

*1.2. The Importance of Accounting Value Measurement*

However, theoretical questions, practical implications, and methodology must go together. Alas, that implies many interdisciplinary, crossing pathways. Yet, when measuring business value, accounting numbers matter the most. Accounting numbers (Aggarwal 2014) are important in order to determine the business performance and other relevant issues. How to interpret and whether the data reported by the business are true or not, is the truly meaningful, essential, question to explore. As much research in the valuation field develops, multiple-based valuation methods (Goumagias 2013; Aggarwal 2014), (i) Free Cash Flow Valuation Model (FCFVM), (ii) Residual Earning Valuation Model (REVM) and (iii) Abnormal Earning Growth Model (AEGM) have emerged being generally accepted as the most popular Valuation Models (VMs) linking with accounting numbers (Farooq and Thyagarajan 2014; Fernandez 2004; Koller et al. 2010). However, these valuation methods if having advantages have also drawbacks with specific difficulties at implementation time. Nevertheless, a valuation model specifies what should be forecasted and further how it is converted to a valuation. Forecasting is at the heart of the process of fundamental analysis and econometrics.

*1.3. Objectives and Research Method of the Study*

It is never being denied that an organization primary goal is supposed to be stated in terms of "economic value" measures, which will assist the firm's internal goals with the maximisation of shareholder value (Copeland et al. 1994; Stern et al. 1995). Hence, the "value" and "the way to show the value" are two equally important subjects to be explored. Both VBM areas and valuation models have seen significant developments in the past few decades. However, there is a lack of research focusing on the relations and the co-effects of these two topics. It seems important to explore the fundamental connection. This paper necessarily aims at analysing the extent to which the use of accounting numbers (Aggarwal 2014) shows the real valuation of business and thus to boost the most efficient VBM systems.

After a literature review section, pointing to the most relevant results, on models (Goumagias 2013) pertinent for our discussion, the implementation section focuses on a quantitative aspect: the retail company, Marks & Spencer (known as M&S) is selected to develop implementations of its financial information along different methods. Because, as pointed by Hannington (2016) on «How to measure and manage your corporate reputation», "when Marks & Spencer's sliding sales was failing to listen to customers effectively, Andersen's questionable business practices revealed a view of the organisation that its customers and stakeholders rejected. They did not want to be associated with a company that had so badly damaged its reputation." Reputation is tied to audits and valuation in a complex way indeed (Beaver 1999; Jones, Comfort, and Hillier 2005).

Thus, the behaviour of M&S customers and that of its shareholders are clearly an interesting study case at this time for emphasizing arguments. The selection of the most effective valuation method(s) in business value creation under the concept of VBM necessarily follows. Tesco being as reputed as M&S, we address limitations and disadvantages and conclude thereafter. In fine, we conclude that valuation models, together with accounting numbers, lead to confidence about a positive answer to the raised question(s).

## 2. Literature Review

This section is divided into four parts: (1) the development of VBM with its advantages and disadvantages in real life; (2) basic assumptions of (i) FCFVM, (ii) REVM, (iii) AEGM and multiple-based method as well as how they are derived and their practicalities (advantages and disadvantages); (3) empirical evidence regarding these valuation methods.

*2.1. Value-Based Management (VBM)*

The evolution of managerial accounting practices can be divided into four stages. The first stage occurs before the 1950s, when managerial accounting practice concerns cost determination and financial control as the primary goal by means of budgeting and cost accounting systems. By the mid-1960s, Anthony (1965) produced a management control framework, through which management control is labelled as the process of obtaining resources and effectively use them to achieve organisation's goals. The third stage began in the mid-1980s with shifting managerial concern from planning and control to the reduction of waste in business procedures (Ittner and Larcker, 2001). "More recently", authors have examined the possibility for firms to provide opportunities for social entrepreneurship and to base business valuation on modern institutional complexity approaches (Cherrier et al. 2018).

It is commonly admitted that every single decision and activity of a company's managers have an effect on the company's value, whereas the lack of integrated process among these decisions and activities often result in inefficient value creation (Plowman 2014). VBM fills in this gap by building up an integrated framework for measuring and managing businesses with the explicit objective of creating superior long-term value for shareholders (Ittner and Larcker 2001).

Martin and Petty (2001) outline several models that firms can use to implement VBM systems, including the free cash flow method, the economic value added/market value added (EVA/MVA) method, and the cash flow return on investment approach. Young and O'Byrne (2000) contend that Economic Value Added (EVA) is the best available metric for measuring a firm's value for four reasons: first, an EVA calculation has great accuracy since it presents the difference between the cost of capital and the return on that capital. Second, EVA can be customised depending on company needs. Furthermore, EVA has overcome drawbacks of other metrics: like the possibility of manipulation of net income and earnings per share, the limitation of merely publicly traded entities calculations, and the absence of the cost of equity and the cost of debt, such as cash flow from operations and cash flow return on investment. The accuracy of EVA lies in its calculation of the cost of debt financing and equity financing since the company is able to calculate EVA either for private entities or for divisions within the company (Young and O'Byrne, 2000).

Nevertheless, Shaked et al. (1997) doubt the practicality of EVA, because of the "superficial simplicity" of EVA, yet labelled this as one of the advantages of the metric. Indeed, a few factors or adjustments have to be determined before EVA is applied. Since they depend on companies' situation, it is difficult to propose universal rules. The process will become distorted if too few factors and adjustments are taken into consideration, whereas it becomes rather complex if too many factors and adjustments are chosen. Moreover, Brewer et al. (1999) argue that the EVA ratio does not necessarily indicate the efficiency of a company. For instance, one company can generate a high EVA but with a low return on investment.

Consequently, other methods to supplement value-based management next to EVA should be considered.

*2.2. How to Measure a Company Value?*

The market value of a company is usually defined as the sum of net debt (the debt which the company holds on its balance sheet as obligations minus the debt which the company nevertheless holds as assets) and the value of equity. Practically, equity is estimated based on its fair value whereas the net debt is reported at book value, usually close to market value. The three main valuations methods (FCFVM, REVM, and AEGM) or so called "models" (Goumagias 2013) of interest here are next introduced.

2.2.1. Free Cash Flow Valuation Method (FCFVM)

Usual cash flow patterns within a company contain the transactions between a firm and its shareholders, for instance, the dividends, internal movements between one component of net financial liabilities and another, for example, an issue or a repayment of debt, and cash flow arising from the operating activities of the company, like net of amounts invested in the operating activities (Penman 2010; Lee 1999). This cash flow is termed as free cash flow (FCF): in FCF theory or model, it is the only cash flow that generates business value. In equity valuation, most analysts forecast FCF to predict entity value (EnV), and then subtract net financial liabilities and non-controlling interest to get the equity value (Demirakos et al. 2004).

Although FCF itself is often used, it does not have a clear and unambiguous definition (Imam et al. 2008). It can be defined as cash flow that is free after the company's operating activities and financial activities (Lee 1999) or, alternatively, as the amount of cash that the operating entity of the company pays to (or receives from) the pool of the company net financial assets (Demirakos et al. 2004). In Penman (2010), there are three methods to calculate FCF from income statement and balance sheet in detail.

Although the Free Cash Flow Valuation Method (FCFVM) is a straight application of the familiar discounted cash flow technique and cash flows are not affected by accounting rules (Copeland et al. 1994), it has some drawbacks. The FCF itself does not measure the value added in the short run. It has a matching problem: the value gained is not matched with the value given up. Moreover, it fails to recognize the value generated which does <u>not</u> involve cash flows, leading to some ambiguity in the valuation.

Additionally, FCFVM has limitation for forecast horizons. According to Lee (1999) and Demirakos et al. (2004), FCFVM works best when expected future post-horizon flows are positive and growing at a constant growth rate (*g*); thus, free cash flow based valuation often requires forecasts for long horizons, particularly if large negative free cash flow is expected in immediate future years. Moreover, it may be difficult to use negative free cash flows as a basis for valuation involving continuing-value terms (Lee 1999; Demirakos et al. 2004).

2.2.2. Residual Earnings Valuation Model (REVM)

Residual earning (RE), also termed as residual income, is defined mathematically as earnings less a cost of capital charge for the book value from a company's perspective. There are different Residual Earnings Valuation Model (REVM) formulae: either for the equity value (EqV) or for the entity value (EnV). For the equity perspective, one has for the residual earnings (*RE*)

$$RE_t = Earn_t - (\rho_E - 1)B_{t-1} \qquad (1)$$

in terms of the return rate on common equity, with the book value $B_t$ obtained according to the balance sheet, at time *t*, while ($\rho_E - 1$) is the required return for common equity (the equity cost of capital); thus, the second term is the Benchmark (*B*) forecast of comprehensive earnings. One observes that there are two drivers for *RE*: the return on common equity (*ROCE*) and the growth in book value. Henceforth, *RE* can be also expressed as follows

$$ROCE_t = \frac{Earn_t}{B_{t-1}} \Rightarrow Earn_t = ROCE_t B_{t-1} \qquad (2)$$

Thus,

$$RE_t = Earn_t - (\rho_E - 1)B_{t-1} = ROCE_t B_{t-1} - (\rho_E - 1)B_{t-1} = (ROCE_t - (\rho_E - 1))B_{t-1} \qquad (3)$$

For the entity perspective, the residual operating income (*ROI*) is

$$ROI_t = OI_t - (\rho_F - 1)NOA_{t-1} \tag{4}$$

where *OI* is the operating income and *NOA* is the net operating asset; $\rho_F$ is the weighted average cost of capital (WACC) (or ($\rho_F$ − 1) is the cost of capital).

The formulae are conditional to the clean surplus relationship (CSR) to be satisfied, which means (Lee et al. 1999; Penman 2001) that

$$B_t = B_{t-1} + Earn_t - d_t \tag{5}$$

where $d_t$ is the dividend paid at time *t*.

Penman (2010) considers the REVM as comprising an anchor (book value) and a premium over book value (made up of the present value of expected future residual earnings). If a firm is expected to earn (more than) a normal rate of return on its capital employed (causing positive residual earnings), the intrinsic value of its equity will equal (exceed) the book value of its equity (Lee, 1999).

Another way to explain REVM (Ohlson 1995; Ohlson 2001; Feltham and Ohlson 1995; Aggarwal 2014) is through "linear information models" (LIMs). These models express the company intrinsic value in terms of earnings and book value, using time series properties of *RE* (and sometimes other accounting items), together with other information (value relevant events for future *RE*), to estimate present value of expected future *RE*.

Ohlson's (1995) LIM assumes that expected future *RE* is asymptotically equal to zero and that book value is approximately equal to intrinsic equity value (under unbiased accounting). The LIM model reads

$$RE_{t+1} = \omega_1 RE_t + v_t + e_{1,t+1} \tag{6}$$

$$v_{t+1} = \gamma_1 v_t + e_{2,t+1} \tag{7}$$

where *v* is some other information, $\omega_1$ and $\gamma_1$ are LIM parameters (non-negative and less than one); $e_1$ and $e_2$ are zero-mean disturbance terms. Combining REVM with LIM, the intrinsic value estimate (Ohlson, 1995) is

$$V_t^E = B_t + \alpha_1 RE_t + \alpha_2 v_t \tag{8}$$

where

$$\alpha_1 = \frac{\omega_1}{\rho_E - \omega_1} \tag{9}$$

$$\alpha_2 = \frac{\rho_E}{(\rho_E - \omega_1)(\rho_E - \gamma_1)} \tag{10}$$

Using Equation (1) and Equation (3), Equation (8) can be rewritten as

$$V_t^E = B_t + \alpha_1 Earn_t - \alpha_1(\rho_E - 1)(B_t - Earn_t + d_t) + \alpha_2 v_t \tag{11}$$

From these formulae, if there is no other information, i.e., $\alpha_2 v_t = 0$, the intrinsic equity value estimate is a weighted average of book value (with weighting $1 - \alpha_1(\rho_E - 1)$) and an ex-dividend earnings multiple (with weighting $\alpha_1(\rho_E - 1)$). Further calculation and analysis show that when $\omega_1 = 0$, *RE* is totally transitory and book value is the dominant explanatory variable: intrinsic value is mainly given by book value (scaled level of earnings). When $\omega_1 = 1$, RE is highly persistent and earnings is the dominant explanatory variable; hence the intrinsic value is largely given by the ex-dividend earnings multiple, or price/earnings (P/E) ratio (equivalent to a scaled change of earnings).

Unlike Ohlson (1995), the Feltham and Ohlson (1995) LIM allows RE to be systematically greater than zero, and its intrinsic value to be systematically greater than the book value (under conservative accounting). Feltham and Ohlson LIM separates the accounting book value into the net financial assets (NFA) and the net operating assets (NOA) where NOA = Operating Assets (OA)—Operating Liabilities (OL) and assume that the RE from NFA is zero. Thus, the relation between entity value $V_t^E$ and equity value $V_t^F$ can be written as

$$V_t^E = V_t^F + NFA_t \tag{12}$$

Feltham and Ohlson LIM model relies on

$$ROI_{t+1} = \omega_0 NOA_t + \omega_1 ROI_t + v_t + e_{1,t+1} \quad (\omega_0 \geq 0, \ 0 \leq \omega_1 < 1) \tag{13}$$

$$NOA_{t+1} = gNOA_t + e_{3,t+1} \quad (g < \rho_F) \tag{14}$$

and also Equation (12) (0 ≤ γ < 1). Using the above information, the intrinsic value estimate (Feltham and Ohlson 1995) is

$$V_t^F = NOA_t + \alpha_1 RE_t + \alpha_2 v_t + \alpha_3 NOA_t \tag{15}$$

where *RE* are still the residual earnings, and

$$\alpha_1 = \frac{\omega_1}{\rho_F - \omega_1} \tag{16}$$

$$\alpha_2 = \frac{\rho_F}{(\rho_F - \omega_1)(\rho_F - \gamma_1)} \tag{17}$$

$$\alpha_3 = \frac{\rho_F \omega_0}{(\rho_F - \omega_1)(\rho_F - g)} \tag{18}$$

Then, using Equation (12)

$$\begin{aligned} V_t^E = V_t^F + NFA_t &= NOA_t + \alpha_1 RE_t + \alpha_2 v_t + \alpha_3 NOA_t + NFA_t \\ &= V_t^E = B_t + \alpha_1 RE_t + \alpha_2 v_t + \alpha_3 NOA_t \end{aligned} \tag{19}$$

Compared with Ohlson (1995) LIM, the additional term $\alpha_3 NOA_t$ in Feltham and Ohlson LIM shows the "conservatism correction": average positive residual earnings are expected from non-recognition and undervaluation of net operating assets. Lee et al. (1999) and Richardson and Tinaikar (2004) figure out that Ohlson (1995) and Feltham and Ohlson (1995) differ in the valuation estimate through effectively combining historical and future valuation statistics, but this is too complicated to be implemented practically. Thus, many new valuation techniques have been invented, but many based on Ohlson (1995) and Feltham and Ohlson (1995).

REVM focuses on value drivers like profitability of investment, growth in investment and strategy. It is economically meaningful and reliable which makes use of the component of value already recognized in the balance sheet (book value). Besides, it uses the properties of accrual accounting, which recognizes value added ahead of cash flows, matches value added to value given up and treats investment as an asset rather than as a cost, giving rise to smoother series of forecasts flows (Lee et al. 1999; Penman 2010).

Because of the smoothing effect of accrual accounting, forecast horizons (prior to a smooth-growth CV term) can be shorter than when free cash flows are used. A significant proportion of value is on the balance sheet, and does not therefore form part of the forecast-flow component, in contrast to FCFVM for which all values are in the forecast. However, in order to implement REVM one needs to understand how accrual accounting works to identify causes for suspicious concern (Penman and Sougiannis 1998; Frankel and Lee 1998).

2.2.3. Abnormal Earnings Growth Model (AEGM)

The Abnormal Earnings Growth Model (AEGM) states the relationship between price, forward earnings and abnormal earnings growth. Penman (2010) defines the abnormal earning growth (AEG) as

$$AEG_t = [Earn_t - Earn_{t-1}] - (\rho_E - 1)[Earn_{t-1} - d_{t-1}] \tag{20}$$

The first term on the right-hand side of Equation (20) is the change in earnings. The second term on the right-hand side (a normal return on time $t-1$ retained earnings) is the extra earnings that the firm would generate at time $t$ if it were to earn a normal return on the $t-1$ reinvested earnings.

AEGM is calculated as follows from

$$V_0^E = \frac{Earn_1}{\rho_E - 1} + \sum_{t=1}^{\infty} \frac{AEG_{t+1}}{(\rho_E - 1)\rho_E^t}, \quad (21)$$

with $AEG_{t+1}$ as in Equation (20)

$$V_0^E = \frac{Earn_1}{\rho_E - 1} + \sum_{t=1}^{T} \frac{AEG_{t+1}}{(\rho_E - 1)\rho_E^t} + \frac{AEG_{T+2}}{(\rho_E - 1)\rho_E^T(\rho_E - g)} \quad (22)$$

In contrast, for the equity perspective, the latter equation with a so called Continuing value (CV) term, the value of free cash flows after the horizon $T$ (Aggarwal 2014) reads in terms of the "abnormal operating income growth" (*AOIG*), and where *OI* is still the operating income,

$$AOIG_{t+1} = (OI_{t+1} - OI_t) - (\rho_F - 1)(OI_t - FCF_t), \quad (23)$$

$$V_0^F = \frac{OI_1}{\rho_F - 1} + \sum_{t=1}^{\infty} \frac{AOIG_{t+1}}{(\rho_F - 1)\rho_F^t}, \quad (24)$$

Below it is shown that AEGM expresses the intrinsic value of equity as the capitalized next-period expected earnings (the value estimate from applying the forward P/E (Price-earning) ratio) plus the present value of the capitalized forecast AEG of subsequent periods (Penman 2010). Indeed,

$$\frac{V_0^E}{Earn_1} = \frac{1}{\rho_E - 1} + \frac{\sum_{t=1}^{\infty} \frac{AEG_{t+1}}{(\rho_E - 1)\rho_E^t}}{Earn_1} \quad (25)$$

If accounting obeys CSR, *AEG* is equal to the first difference of *RE* (Ohlson 2005):

$$\begin{aligned} AEG_{t+1} &= (Earn_{t+1} - Earn_t) - (\rho_E - 1)(Earn_t - d_t) \\ &= (Earn_{t+1} - Earn_t) - (\rho_E - 1)(B_t - B_{t-1}) \\ &= (Earn_{t+1} - (\rho_E - 1)B_t)) - (Earn_t - (\rho_E - 1)B_{t-1})) \\ &= RE_{t+1} - RE_t. \end{aligned} \quad (26)$$

One can easily write the equivalent equations for the equity perspective.

Notably, AEGM is more complicated to calculate than FCFVM and REVM, but focuses more on P/E ratio and future earnings. Yet, analysts use AEGM, P/E ratio and future earnings to measure a company's earnings per share (EPS) and EPS growth. Nevertheless, one needs to consider to what extent dividend policy irrelevancy can work in these models. Under AEGM, CSR is not required. Therefore, AEGM can be applied on a per-share basis, compared with REVM that cannot be used if there are any share transactions (Ohlson 2005). Additionally**,** AEGM can be more error free if the error (such as improper valuation of assets and liabilities) in two consecutive balance sheets is the same, since one can still rely on income statement for valuation. However, AEGM does not include any explicit reference to the balance sheet—which is often an important driver of earnings growth (Ohlson 2005; Penman 2010).

2.2.4. Multiple-Based Valuation Methods

Multiple-based valuation methods value an entity through comparative companies. The value of the entity is calculated as the value driver (such as sales and net profit) multiply by comparative firms' multiples (from their mean, median or another measure of central tendency) (Penman 2010).

The process of multiple-based valuation is straightforward. First, one has to identify comparable firms that have similar characteristics to the target firm. This could be a set of firms or one specific firm. Then, feasible measures need to be identified for the comparable firms in their financial statements (such as earnings, book value, sales) and calculate multiples of those measures at which the firms trade, such as EnV/Sales (Entity value over sales). Finally, one has to apply these multiples to the corresponding measures (mean or median or another measure of central tendency) for the target, i.e., to estimate that firm value (Alford 1992; Liu, Nissim and Thomas 2002).

Many scholars try to find ways to obtain accurately comparable firms for valuing a specific firm. Alford (1992) examines the relative accuracy of pricing-earnings multiples when comparable firms are selected, according to industry, size, leverage, and earnings growth. Using price-scaled absolute prediction error, he finds that valuation errors decline when the industry classification definition is narrowed to two or three-digit SIC codes, but no further improvement when a four-digit classification is used. Moreover, further controls for firm size, leverage, and earnings growth do not reduce valuation errors, after controlling for industry membership.

Liu et al. (2002) complement Alford (1992) on exploring how the accuracy of multiples-based valuation methods varies across dimensions, in particular across different value drivers (e.g., book value, earnings, forecast earnings, cash flows) and across different sets of comparable firms (all firms and firms from the same industry only). Using harmonic mean or complex regression-based approach with non-zero intercept, and a signed scaled pricing error instead of absolute ones as in Alford (1992), they find that in those multiples, forward earnings multiples perform best in estimation, followed by historical earnings, cash flow multiples, and book value of equity; sales perform the worst. Contrary to the generally accepted view that different industries have different suitable multiples, Liu et al. (2002) find that the ranking here above is consistent for almost all industries which were examined.

In the same year, Bhojraj and Lee (2002) propose that for selecting a comparably based multiple in a firm valuation, comparable firms can be identified by a measure called "warranted multiple", from multivariate regression combining the firm-specific characteristics (e.g., profitability, growth) and the firm sector's average. Using two valuation multiples, entity value to sales and price-to-book ratio, they argue that the firms that are most comparable to the specific firm are those with the warranted multiples closest to that of firm. They find that relative to multiples for firms with a simpler (and more traditional) industry matching used, the explanatory power for valuation multiples is substantially higher when multiples from warranted-multiple matching are used.

Although such multiple-based valuations are easy to understand and calculate, they present disadvantages. First, finding comparable firms that precisely match, is difficult and always subject to caution. Indeed, the multiples may be largely affected by accounting policy choices, relative financial risk, and some subjective judgments on different companies, causing confusion at comparison time. Moreover, when calculating target multiples, the mean value of the multiples may be distorted by extreme values, outliers (thus it might be advisable to use median or harmonic mean for better arguing). Additionally, multiple-based valuation assumes a linear relationship between the entity value and value drivers, which may not often strictly hold in practice (Penman 2010).

*2.3. Further Discussion on Valuation Techniques*

Theoretically, REVM should give the same estimate of the intrinsic value of equity as FCFVM and AEGM, regardless of the accounting recognition and measurement conventions applied, provided that forecasts of dividends (net of equity issues), balance sheet values and earnings are consistent with each other in accordance with CSR (Lundholm and O'Keefe 2001a). In practice, this premise will not always hold, causing some models to give better estimates than others, resulting in discrepancy of valuation figures (Demirakos et al. 2004).

Francis et al. (2000) (FOO) practically compared the relative performance of FCFVM and REVM in producing value estimates, that is, how close the estimates are compared to the observed price. FOO compare these models through statistics of signed prediction errors (bias to observed prices), absolute prediction errors (inaccuracy) and association with observed prices (R-squared statistics in a univariate regression of the observed price on the value estimate). They conclude that REVM generally perform more accurately than FCFVM (and the Farrell (1985) Dividend Discount Model (DDM)). Penman and Sougiannis (1998) (PS) carried on a similar analysis but in a more complicated way regarding assumptions of CV terms and realisations of payoff (accounting policy differences). Their conclusion is consistent with FOO: REVM's mean valuation error is smaller than DDM and FCFVM.

However, both FOO and PS's findings are criticised by Lundholm and O'Keefe (LO1a) (2001a) and Lundholm and O'Keefe (2001b) (LO1b) who claim that the same intrinsic value should be obtained if implemented consistently. LO1a argue that FOO and PS use inconsistent implementation due to either data inconsistency (such as violation of CSR) or assumption inconsistencies (such as same growth rate for RE and dividend). However, Penman (2001) disagrees with LO1a and argues that accounting-based valuation models (e.g., REVM) could dominate cash-based approaches (e.g., FCFVM) if making finite horizon forecasts in the valuation. However, LO1b still insist that if carefully done and make full use of financial statements, the results of intrinsic value from FCFVM and REVM should be the same.

Lee et al. (1999) agree with FOO's opinion that REVM performs well by testing multiples. They use time-series regression co-integrating price and intrinsic value to make price and intrinsic value long-term convergent, and compare the performance of estimates of intrinsic value for the Dow 30 stocks. From 1963 to 1996, ratios based on REVM had statistically much more reliable predictive power than traditional market multiples, like book-to-price, earning-to-price and dividend-to-price ratios.

Frankel and Lee (1998) (FL) use a different approach from FOO, PS, and Lee et al. (1999) to show that REVM works well. They implement REVM for a large sample of firms using analysts' forecasts and obtain a set of valuations (V). Then FL use the value to price ratio (V/P) to predict stock return and compare its performance with book-to-price ratio via post-valuation-date returns. The finding is: V/P ratio is a better predictor of future returns, particularly over longer horizons.

Despite the valuation methods mentioned here above, several practitioners suggest other valuation approaches. For instance, Burgstahler and Dichev (1997) develop an option-style equity valuation which combines company's capitalised expected earnings (going concern concept satisfied) with value of firm's adaptation option (conversion of firm's resource to more effective use). They argue that the equity value has a convex function with both expected earnings and book value of a specific company, instead of a simple additive relation of the two components suggested by Ohlson (1995). However, this valuation model is too complex for determining the coefficient of the function.

Therefore, following such considerations, it can be concluded that only useful accounting numbers can produce a meaningful and accurate business valuation. Thus, FCFVM, REVM, AEGM and Multiple-based valuation method can be generally accepted as the most commonly used valuation methods linking with accounting numbers. However, these methods have their own benefits and drawbacks. The most reliable valuation method out of the four seems to be REVM.

## 3. Implementation

### 3.1. Overview

In order to implement, analysis and comparison of these valuation methods on a quantitative basis, we selected a well-known company, Marks & Spencer (M&S). Sources of implementation are from M&S's official website, its annual reports, and several analyst reports—see the reference list.

N.B. Brea-Solís et al. (2015) in "*Business Model Evaluation: Quantifying Walmart's Sources of Advantage*", obviously studying Walmart (rather than Mark & Spencer, as in our case), have shown that the effectiveness of a particular business model depends not only on its design (its levers and how they relate to one another) but also, most importantly, on its implementation.

M&S is one of the major retailers in the UK. It sells outstanding quality food with stylish, high quality, great value clothing and home products from global suppliers. It has approximately 78,000 employees and 700 stores in the UK and operates in 42 countries all over the world and is expanding its overseas business (Marks & Spencer 2016). In recent annual report, it shows GBP 9.5 billion revenue and GBP 404.4 million net profit, which correspondingly has a growth of 2.7% and -16% compared to last year.

*3.2. Valuation Implementation*

Thus, in line with the above, two essential steps need to be followed to arrive at accurate value estimates (Penman 2010). Original financial statements need to be first reformulated then forecasted, for facilitating flow-based calculations. All valuation methods are implemented at the entity level. The entity-level value is then adjusted for net financial assets or liabilities, minority interest and other relevant factors to present an estimate of the intrinsic value of equity. Divided by number of shares in issue, the final result is the intrinsic value per share.

*3.3. Reformulation of Financial Statements*

Balance sheet and income statement items can be classified into operating items and financial items. Reformulation of income statement and financial statement allows one to separate operating and financial items while keeping the total amount unchanged, as the company value is mostly determined by the operating items (the underlying business), since financial items can be "noisy" for the forecast.

*3.4. Forecast of Financial Statements*

The forecast of key inputs is mainly based on M&S's website information, its annual reports, relative analyst reports and reasonable estimate. The data used in these sources or a reasonable estimate are accepted as reliable after comparison and analysis. The forecast horizon is intended to be 5 years long, from 2017 to 2021. Sales growth is forecasted at 2% stably till infinity. Profit margin stays at the 9% level according to company's cost control ability. Most balance sheet items are assumed to be proportional to net operating asset. According to sales growth, the net operating asset is growing at 2% since 2017. Keeping other comprehensive income (OCI) proportional to sales for simplicity, the tax rate is assumed to be 28%. We keep financial items and non-controlling interest proportional for simplicity (or because lacking relevant information). Number of shares in issue is 1635.90 million.

The results of the calculation process (valuation date as at 31/3/2016) for each method can be seen in the tables, where the lines followed the reasoning and the equations here above given. For FCFVM, only Methods 1 and 2 (Penman 2010) are used in implementation; Method 3 is complicated and unfeasible in this case.

*3.5. Comparing these Valuation Methods*

From the above, it can be seen that the intrinsic share price and the intrinsic value of Marks & Spencer are the same (3.08 pound and 6945.31 million), although they are obtained from different valuation approaches of FCFVM, REVM and AEGM. Table 1 summarizes the three valuation methods bringing the same entity valuation figure together. Through the tables, it is thus demonstrated that the three models, FCFVM, REVM, and AEGM, can reasonably be said to give identical value estimates—see also Lundholm and O'Keefe (2001a).

**Table 1.** Consistent Implementation of Free Cash Flow Valuation Model (FCFVM), Residual Earning Valuation Model (REVM) and Abnormal Earning Growth Model (AEGM).

|  | 2016 | 2017E | 2018E | 2019E | 2020E | 2021E |
|---|---|---|---|---|---|---|
| Net operating income (including other comprehensive income items) | 483.20 | 446.84 | 455.77 | 464.89 | 474.19 | 483.67 |
| Inventories | | 799.90 | 815.90 | 832.22 | 848.86 | 865.84 | 883.15 |
| Trade and other receivables | | 321.10 | 327.52 | 334.07 | 340.75 | 347.57 | 354.52 |
| Current tax receivable | | 1.60 | 1.63 | 1.66 | 1.70 | 1.73 | 1.77 |
| Trade and other payables | | −1617.70 | −1650.05 | −1683.06 | −1716.72 | −1751.0 | −1786.07 |
| Current tax liabilities | | −75.20 | −76.70 | −78.24 | −79.80 | −81.40 | −83.03 |
| Working capital | | −570.30 | −505 | −593.35 | −641.14 | −617.31 | −629.66 |
| PPE and intangible assets | | 5829.90 | 5946.50 | 6065.43 | 6186.74 | 6310.47 | 6436.68 |
| Other net operating assets | | −67.60 | −68.95 | −70.33 | −71.74 | −73.17 | −74.64 |
| Net operating assets | | 5192 | 5372.55 | 5401.75 | 5473.86 | 5619.99 | 5732.38 |
| *Summary of movement in property, plant and equipment (PPE) (or "tangible assets") and other intangible assets:* | | | | | | | |
| Brought forward | | 5889.30 | 5829.90 | 5946.50 | 6065.43 | 6186.74 | 6310.47 |
| Additions less disposals | | 503.40 | 513.47 | 523.74 | 534.21 | 544.90 | 555.79 |
| Depreciation and amortisation | | −562.80 | 396.87 | 404.81 | 412.90 | 421.16 | 429.58 |
| Carried forward | | 5829.30 | 5946.50 | 6065.43 | 6186.74 | 6310.47 | 6436.68 |
| *Estimate of free cash flow (in million GBP)* | | | | | | | |
| **Method 2** | | | | | | | |
| Net operating income (including other comprehensive income items) | | | 446.84 | 455.77 | 464.89 | 474.19 | 483.67 |
| Minus: increase in receivables, etc | | | −6.42 | −6.55 | −6.68 | −6.82 | −6.95 |
| Minus: increase in inventories, etc. | | | −16.00 | −16.32 | −16.64 | −16.98 | −17.31 |
| Plus: increase in payables, etc | | | 32.35 | 33.01 | 33.66 | 34.33 | 35.02 |
| Minus: increase in other net operating assets | | | 1.35 | 1.38 | 1.41 | 1.43 | 1.47 |
| Plus: depreciation and amortisation | | | 396.87 | 404.81 | 412.90 | 421.16 | 429.58 |
| Minus: additions (net of disposals) etc. to PPE and intangible assets | | | −818.86 | −835.23 | −851.94 | −868.98 | −886.36 |
| Estimate of free cash flow | | | 339.77 | 346.55 | 353.49 | 360.56 | 367.77 |
| **Method 1** | | | | | | | |
| Net operating income (including other comprehensive income items) | | | 446.84 | 455.77 | 464.89 | 474.19 | 483.67 |
| Minus: Change in net operating assets | | | −107.07 | −109.22 | −111.40 | −113.63 | −115.90 |
| Estimate of free cash flow | | | 339.77 | 346.55 | 353.49 | 360.56 | 367.77 |
| *FCF Entity Valuation (million GBP)* | | | | | | | |
| Free cash flow (FCF) | | | 339.77 | 346.55 | 353.49 | 360.56 | 367.77 |
| Discount rate @ 7% | | | 1.07 | 1.14 | 1.23 | 1.31 | 1.40 |
| Present value of free cash flows | | | 317.54 | 303.99 | 287.39 | 275.24 | 262.69 |
| Total present value to 2021 | | 1446.85 | | | | | |
| Continuing value (CV) (g = 2%) | | | CV = 367.77 * (1 + 2%)/(7% − 2%) => | | | | 7501.51 |
| Present value (PV) of CV | | 5358.22 | | | | | |
| Entity value (EnV) | | 6945.31 | | | | | |
| *REVM Entity Valuation (million GBP)* | | | | | | | |
| Net operating income (including other comprehensive income items) | | | 446.84 | 455.77 | 464.89 | 474.19 | 483.67 |
| Net operating assets | | 5353.70 | 5460.77 | 5569.99 | 5681.39 | 5795.02 | 5910.92 |
| Capital charge @7% | | | 374.76 | 382.25 | 389.90 | 397.70 | 405.65 |
| Residual Operating Income (ROI) | | | 72.08 | 73.52 | 74.99 | 76.49 | 78.02 |
| Discount rate @ 7% | | | 1.07 | 1.14 | 1.23 | 1.31 | 1.40 |
| PV of ROI | | | 67.36 | 64.50 | 60.97 | 58.39 | 55.73 |
| Total PV of ROI to 2021 | | 306.95 | | | | | |
| CV of ROI | | | CV = 78.02 * (1 + 2%)/(7% − 2%) => | | | | 1591.61 |
| PV of CV of ROI | | 1284.66 | | | | | |
| Entity value (EnV) | | 6945.31 | | | | | |
| *AEGM Entity Valuation (million GBP)* | | | | | | | |
| Net operating income (including other comprehensive income items) | | | 446.84 | 455.77 | 464.89 | 474.19 | 483.67 |
| Free cash flow (FCF) | | | 339.77 | 346.55 | 353.49 | 360.56 | 367.77 |
| Prior year reinvested | | | | 107.07 | 109.22 | 111.40 | 113.63 |
| Normal change in earnings@7% | | | | 6.26 | 6.37 | 6.53 | 6.68 |
| Change in operating income | | | | 8.93 | 9.12 | 9.3 | 9.48 |
| Abnormal operating income growth (AOIG) | | | | 2.67 | 2.75 | 2.77 | 2.8 |
| Capitalised next period AOIG@7% | | | 44.82 | 45.71 | 46.63 | 47.56 | |
| Discount rate @ 7% | | | 1.07 | 1.14 | 1.23 | 1.31 | |
| PV of cap AOIG | | | 41.89 | 40.10 | 37.91 | 36.31 | |
| Total PV of cap AOIG to 2021 | | 180.23 | | | | | |
| CV of cap AOIG | | CV = 3.84 * (1 + 2%)/[(7% − 2%) * 7%] => | | | | | 1119.80 |
| PV of CV of cap AOIG | | 854.29 | | | | | |

| Capitalised operating income for 2017 | 7127.51 |
|---|---|
| Entity value (EnV) | 6945.31 |

*3.6. Sensitivity Analysis*

The sensitivity analysis assesses the method robustness in its assumptions when changing one variable input in the model and holding constant the others. Take FCFVM as an example: percentage figures such as terminal sales growth rate and the Weighted Average Cost of Capital (WACC) are expected to be very "sensitive". Table 2 shows a quantitative comparison and the relationship between these two factors if they are increased or decreased by 1%, keeping other conditions constant. It is seen that both entity and equity value of M&S change substantially, when the WACC or the growth rate change by only 1%. REVM and AEGM lead to an outcome similar to that of FCFVM. Interestingly, it can be seen that the company value is positively related to sales growth but is negatively related to WACC.

**Table 2.** Sensitivity analysis (in million GBP).

|  | *g* = 2% | | | WACC = 7% | | |
| --- | --- | --- | --- | --- | --- | --- |
|  | WACC = 6% | WACC = 7% | WACC = 8% | *g* = 1% | *g* = 2% | *g* = 3% |
| PV (2017–2021) | 1785.10 | 1736.93 | 1690.77 | 1736.93 | 1736.93 | 1736.93 |
| CV (Continuing Value) | 8417.43 | 6425.10 | 5110.92 | 5301.76 | 6425.10 | 8110.12 |
| Entity value (EnV) | 10202.54 | 8162.03 | 6801.69 | 7038.69 | 8162.03 | 9847.04 |
| EnV change | 25.00% | 0% | −16.67% | −13.76% | 0% | 20.64% |
| Equity value (EqV) | 8428.74 | 6388.23 | 5027.89 | 5264.89 | 6388.23 | 8073.24 |
| EqV change | 31.94% | 0% | −21.29% | −17.58% | 0% | 26.38% |

*3.7. Multiple-Based Method Valuation*

Based on M&S's industry and business pattern (size, growth prospect), two comparable companies, Tesco and Sainsbury's, are chosen for the multiple-based valuation. Both companies are major retailing companies in the UK and have a significant market share and customer group (Tesco 2016; Sainsbury's 2016).

Two value drivers are chosen, EnV-to-EBIT (entity value over earnings before interests and taxes) and EnV-to-sales. As the literature review suggested, median or harmonic mean for multiples can be used in this case when there are extreme outliers in the multiples. Table 3 shows the detailed calculation process of multiple-based valuation for each value driver. The final reported result of entity value and share price are the averages of the four entity values and share prices.

**Table 3.** Results of multiple-based valuation: left column calculation using median, while right column is using the harmonic mean, for either value driver.

|  | EnV/EBIT | | EnV/Sales | |
| --- | --- | --- | --- | --- |
| Companies | **Tesco** | **Sainsbury's** | **Tesco** | **Sainsbury's** |
| Multiples | 10.6 | 11.0 | 1.0 | 0.6 |
| Marks & Spencer's multiple (median) | 10.8 | - | 0.8 | - |
| Marks & Spencer's multiple (harmonic mean) | - | 10.53 | - | 0.77 |
| EBIT | 746.50 | 746.50 | - | - |
| Sales | - | - | 9934.30 | 9934.30 |
| Entity Value | 8062.20 | 7860.65 | 7947.44 | 7649.41 |
| Net financial liabilities | −1762.40 | −1762.40 | −1762.40 | −1762.40 |
| Non-controlling interest | −11.40 | −11.40 | −11.40 | −11.40 |
| Intrinsic value of equity | 6288.40 | 6086.85 | 6173.64 | 5875.61 |
| Shares outstanding | 1605.51 | 1605.51 | 1605.51 | 1605.51 |
| Intrinsic value per share | 3.92 | 3.79 | 3.85 | 3.66 |
| Average entity value | 7879.92 | | | |
| Average equity value | 6106.13 | | | |
| Average share price | 3.80 | | | |

*3.8. Brief Conclusion on Implementation*

For better emphasis, some discussion of the methods is already presented in the following subsections; in particular, the difficulties and limitations of the implementation are outlined.

3.8.1. Best Model for Value-Based Management

From the implementation section, it can be observed that FCFVM, REVM and AEGM produce similar estimates regarding a business' value, in agreement with the theoretical hypotheses as previously presented in the literature review. However, from a large body of resources, we have shown that FCFVM, if compared with REVM and AEGM, is the most effective valuation model. Being different from what has been illustrated in Section 2.1, i.e., EVA is widely believed to be most effective method to conduct value-based management, the free cash flow valuation method has shown its superiority relating to VBM. Let us discuss such a consideration on different levels.

3.8.2. The Consistency of the Fundamental Assumption

Free cash flow refers to the cash remaining from all cash expenses and operating investment for a company's daily operation. It is the hard cash that the company's claim holders, especially the shareholder, can rely upon for their own interest (Young and O'Byrne 2000; Damodaran 2012; Bragg 2002). Thus, the concept of free cash flow plays the foundation role of value-based management since the core assumption of VBM is to maximise the shareholders' value. As asserted by Martin et al. (2009), free cash flow is supposed to be placed at the heart of any attempt to learn how management contributes to a company's value.

More importantly, free cash flow not only signifies the available amount distributed to a company's investors, also it represents the core determinant of the company's value (Martin et al. 2009). These determinants of value, often labelled as "value driver", are a particular kind of activity or organisational focus that strengthen the perceived value of the company's product or service in the customers' perception and therefore creates value for the company (Miller 2010). It includes advanced technology, sales, sales growth, operating profit margins, asset-to-sales relationships, reliability, reputation and cash taxes, etc. (Martin et al. 2009; Miller 2010). Consequently, management is supposed to focus on the strategies that relates with "value driver" and thus create value, which goes parallel with the basic assumption of free cash flow.

3.8.3. Calculation Concordance

Moreover, the calculation method of free cash flow also matches the concept of value-base management. Free cash flow methods use the weighted average cost of debt and equity to discount free cash flows (Sofat and Hiro 2016). It is equal to the cash flow from operations less any incremental investments in working capital and capital expenditures (Martin et al. 2009). The present value of net operating cash flows plus the present value of terminal value equals the economic value. According to Sofat and Hiro (2016), the business growth curve is expected to be steep during the planning horizon while cash flows will remain constant or grow at a low rate afterwards, because of the competition with rivals. Moreover, an estimate of terminal cash flows also needs to be done; it can reflect the value of post-planning cash flows when one is calculating the business value over a planning horizon.

3.8.4. Suitable Scenario for VBM

Different methods may suit in different situations instead of VBM; free cash flow method seems to be the most reliable. Free cash flow considers the earnings per share or return on invested capital as adequate criteria of value creation and this has made the model to be capable of evaluating if the management of a firm is creating value or not (Martin et al. 2001). Other methods may deal with lesser information in different cases. For instance, AEGM is used for the company who is not paying dividends but required to value; REVM can be used for companies with stable dividends and leverage; EVA suits the situation where comparisons need to be made on an annual basis and as a performance measure at the same time (Farooq and Thyagarajan 2014). Thereby one concurs that EVA can serve as the benchmark.

3.8.5. Limitations and Difficulties of Practical Implementation

Of course, not every possible calculation method, so far proposed in the literature, has been examined in the implementation part of this paper. Instead, we have chosen the most common and comparable methods in order to reach a set of conclusions.

Moreover, several difficulties in the implementation process are mainly from reformulation, forecast and calculation—some of which having already been discussed in the literature review.

In the reformulation stage, the most challenging part is how to separate items that are partly operating and partly financing. Because the consolidated income statement and balance sheet is relatively simple, several specific items contain "too much information", but they do not specify which part is operating and which part is financing. For instance, in M&S's financial statements, accounts receivable, accounts payable, prepayments and accrual expenses may contain interest-bearing items belonging to financial measures. Deferred income taxes and taxes payable also suffer from this ambiguity. A priori, M&S had seemed to be an interesting and moreover timely case to analyze and to study in order to emphasize pro and cons arguments. This hypothesis is confirmed a posteriori.

The most controversial part seems to be how to determine the items beside sales growth and forecast horizon, when forecasting a reformulated income statement and balance sheet. In this case, the proportional assumption is used for simplicity, but actually most of the items cannot change proportionally because of the company strategy or/and macroeconomic factors. For instance, if a company decides to initiate the strategy of delaying payment, accounts payable will be much larger than the proportional assumption.

The final problem is how reliable analyst reports are. Although analysts' forecast figures are different to a great extent, the figures are somewhat too close, resulted from the herding effect due to which the figures may not be reliable when only one analyst really determines the percentage and others just slightly change a little bit according to the instinct. If information in the analyst reports is inaccurate or even manipulated, the final valuation results will be meaningless. This is why we may suggest to complement the valuation methods with a Benford's Law (Benford 1938; Shi et al. 2018) treatment at first. Moreover, information about sales growth and WACC can be easily found, but those figures are quite sensitive as it has been shown in Section 3.6, and in Table 2. Additionally, it is inevitable to have some forecast bias and rounding errors due to many inter te figures in the annual reports or analyst reports.

4. Conclusions and Suggestions for Future Work

The finance world is no longer merely happy about the numbers. Value based management has been attracting increasing attention. It can be best understood as a "relationship between a value creation mind set and management processes and systems which are required to change that mind set into action" (Sofat and Hiro 2016: 278). It is about value measuring at its most beating heart.

There are three dimensions involved in the concept of value-based management: value creation, value maximisation and value measurement. By applying value-based management, staff inside a corporation, from the executives to the employees, is aware of the importance of value maximisation. They should therefore be able to identify the drivers of value, such as advanced technology, sales, sales growth, operating profit margins, asset-to-sales relationships, reliability, reputation and cash taxes etc. Moreover, they should be able to determine if there is a possibility for certain drivers to be improved in order to maximize the shareholders' business value (Towsend et al. 2019).

This paper has chosen the "measurement through model" dimension of value-based management as the specific pathway for practical considerations. Recall that the reasons are threefold. First, accounting numbers play rather important role in finance world while it only counts once being presented properly. Second, there are various different methods available for people to choose when measuring an organization financial condition, each of them having specific advantages and disadvantages. Finally, and most importantly, there seems to be a lack of research on the relation between the valuation methods and value-based management, i.e., which valuation method suits best for value-based management. Here, we hopefully propose a coherent line of discussions and indubitable conclusions, apparently unavailable in the literature according to our points of view, because there is a need for coherent pathways connecting different disciplines or points of views.

We have answered the question in three steps. First, the reviewed literature, on major valuation methods, including free cash flow valuation method, residual earnings valuation method, abnormal earning growth model, and multiple-based valuation method, has brought forward knowledge, but also has made questions more specific. After reviewing these methods, we decided to choose four methods for illustrating a practical implementation; the second step of this research was both theoretical and methodological.

Coming to the implementation part, in order to complement the literature review and compare these valuation approaches, a well-known retail company Marks & Spencer (M&S) has been chosen to estimate its business value by using its financial statements and relevant analyst reports after reformulation, leading to forecast. The five-year financial information has been chosen to implement different calculation method with reformulation of financial statement at first and different implementation process accordingly. The result is that FCFVM, REVM and AEGM produce the same estimate, which is in agreement ("fortunately", one must admit) with theoretical findings in the literature.

Nevertheless, one has to be aware that some limitations do exist on conclusions when using any of these four valuation approaches. One can argue that we compare too few companies, and add that too few value drivers are used, leading to a likely "not universal" result. From the perspective of FCFVM, REVM and AEGM, those percentage figures (such as sales growth rate, profit margin, WACC) are indeed very sensitive. Essentially, there are several assumptions used either to simplify the forecast or to calculate REVM; one might debate on whether they are (un)realistic. However, we feel some confidence that both from the literature review and from the practical implementation case study, FCF seems a reliable model among these valuation methods, with much advantage, due to its simplicity, accuracy and flexibility.

**Author Contributions:** Conceptualization, H.L. and M.A.; methodology, H.L. and M.A.; software, N.A.; validation, H.L. and M.A.; formal analysis, H.L. and M.A.; investigation, H.L.; resources, N.A.; data curation, H.L. and M.A.; writing—original draft preparation, H.L.; writing—review and editing, M.A.; visualization, H.L. and M.A.; supervision, M.A.; project administration, M.A.; funding acquisition, N.A. All authors have read and agreed to the published version of the manuscript.

**Funding:** This research received no external funding

**Acknowledgments:** This paper is based on a dissertation by H.L. to the School of Business, University of Leicester, with M.A. as supervisor.

**Conflicts of Interest:** The authors declare to have no conflict of interest.

**References**

1. Aggarwal, N. Equity Valuation Using Accounting Numbers. *Int. Res. J. Commer. Arts Sci.* **2014**, *5*, 153–170.
2. Beaver, G. Competitive advantage, corporate governance and reputation management: The case of Marks & Spencer. *J. Commun. Manag.* **1999**, *4*, 185–196.
3. Benford, F. The law of anomalous numbers. *Proc. Am. Philos. Soc.* **1938**, *78*, 551–572.
4. Bhojraj, S.; Lee, C.M.C. Who Is My Peer? A Valuation-Based Approach to the Selection of Comparable Firms. *J. Acc. Res.* **2002**, *40*, 407–439, doi:10.1111/1475-679X.00054.
5. Bragg, S.M. *Business Ratios and Formulas: A Comprehensive Guide*; John Wiley & Sons, Inc.: Hoboken, NJ, USA, 2002. Available online: http://www.vnseameo.org/ndbmai/BRF.pdf (accessed on 20 June 2016).
6. Brea-Solís, H.; Casadesus-Masanell, R.; Grifell-Tatjé, E. Business Model Evaluation: Quantifying Walmart's Sources of Advantage. *Strateg. Entrep. J.* **2015**, *9*, 12–33, doi:10.1002/sej.1190.

7. Brewer, P.C.; Chandra, G.; Hock, C.A. Economic Value Added (EVA): Its Uses and Limitations. *S.A.M Adv. Manag. J.* **1999**, *64*, 4–11.
8. Burgstahler, D.C.; Dichev, L.D. Earnings, Adaptation and Equity Value. *Account. Rev.* **1997**, *72*, 187–215.
9. Chen, S.; Dodd, J. Economic Value Added: An Empirical Examination of a New Corporate Performance Measure. *J. Manag. Issues* **1997**, *9*, 319–333.
10. Cherrier, H.; Goswami, P.; Ray, S. Social entrepreneurship: Creating value in the context of institutional complexity. *J. Bus. Res.* **2018**, *86*, 245–258.
11. Copeland, T.; Koller, T.; Murrin, J. *Valuation: Measuring and Managing the Value of Companies*, John Wiley & Sons: Hoboken, NJ, USA, 1994
12. Damodaran, A. *Investment Valuation: Tools and Techniques for Determining the Value of Any Asset*; John Wiley & Sons, Inc: Hoboken, NJ, USA, 2010.
13. Demirakos, E.G.; Strong, N.C.; Walker, M. What Valuation Models Do Analysts Use? *Account. Horiz.* **2012**, *18*, 221–240, doi: 10.2308/acch.2004.18.4.221.
14. Farrell, J.L., Jr. The dividend discount model: A primer. *Financ. Anal. J.* **1985**, *41*, 16–25.
15. Farooq, M.S.; Thyagarajan, V. Valuation of Firm: Methods and Practices. An Evaluation. *Int. J. Res. Bus. Manag.* **2014**, *2*, 7–14.
16. Feltham, G.A.; Ohlson, J.A. Valuation and Clean Surplus Accounting for Operating and Financial Activities. *Contemp. Account. Res.* **1995**, *11*, 689–731.
17. Francis, J.; Olsson, P.; Oswald, D.R. Comparing the Accuracy and Explainability of Dividend, Free Cash Flow, and Abnormal Earnings Equity Value Estimates. *J. Account. Res.* **2000**, *38*, 45–70.
18. Frankel, R.; Lee, C.M.C. Accounting Valuation, Market Expectation, and Cross-sectional Stock Returns. *J. Account. Econ.* **1998**, *25*, 283–319.
19. Global Data. *Marks and Spencer Group plc (MKS)—Financial and Strategic SWOT Analysis Review*; Global Data: London, UK, 2016.
20. Goumagias, N. Equity Valuation Using Accounting Numbers. M.Sc. in Finance Thesis, Lancaster University, Lancaster, UK, 2013
21. Hannington, T. *How to Measure and Manage Your Corporate Reputation*; Routledge: Abingdon, UK, 2016.
22. Imam, S.; Barker, R.; Clubb, C. The Use of Valuation Models by UK Investment Analysts. *Eur. Account. Rev.* **2008**, *17*, 503–535, doi:10.1080/09638180802016650.
23. Jones, P.; Comfort, D.; Hillier, D. Corporate social responsibility as a means of marketing to and communicating with customers within stores: A case study of UK food retailers. *Manag. Res. News* **2005**, *28*, 47–56.
24. Koller, T.; Goedhart, M.; Wessels, D. *Valuation: Measuring and Managing the Value of Companies,* 5th ed.; John Wiley & Sons, Inc.: Hoboken, NJ, USA, 2010. Available online: https://samples.sainsburysebooks.co.uk/9780470889947_sample_411706.pdf (accessed on 20 June 2016).
25. Lee, C.M.C. Accounting-Based Valuation: Impact on Business Practices and Research. *Account. Horiz.* **1999**, *13*, 413–425, doi:10.2308/acch.1999.13.4.413.
26. Lee, C.M.C.; Myers, J.; Swaminathan, B. What Is the Intrinsic Value of the Dow? *J. Financ.* **1999**, *54*, 1693–1741, doi:10.1111/0022-1082.00164
27. Liberum Capital. *Marks & Spencer, Are We There Yet?* Liberum Capital Ltd.: London, UK, 2012.
28. Liu, J.; Nissim, D.; Thomas, J. Equity Valuation Using Multiples. *J. Account. Res.* **2002**, *40*, 135–172, doi:10.1111/1475-679X.00042.
29. Lundholm, R.; O'Keefe, T.B. Reconciling Value Estimates from the discounted Cash Flow Model and the Residual Income Model. *Contemp. Account. Res.* **2001**, *18*, 311–335, doi:10.1506/W13B-K4BT-455N-TTR2.
30. Marks & Spencer. *Marks and Spencer Annual Report*; Marks and Spencer: London, UK; 2012.
31. Marks & Spencer. *Marks and Spencer Annual Report*; Marks and Spencer: London, UK; 2013.
32. Marks & Spencer. *Marks and Spencer Annual Report*; Marks and Spencer: London, UK; 2014.
33. Marks & Spencer. *Marks and Spencer Annual Report*; Marks and Spencer: London, UK; 2015.
34. Marks & Spencer. Marks and Spencer about Us Company Overview. Available online: https://corporate.marksandspencer.com/aboutus (accessed on 20 June 2016).
35. Martin, J.D.; Petty, J.W. *Value-Based Management: The Corporate Response to the Shareholder Revolution*; Oxford University Press: England, UK, 2001. Available online: http://oxfordindex.oup.com/view/10.1093/acprof:oso/9780195340389.003.0004 (accessed on 26 July 2016).

36. Miller, W.D. *Value Maps: Valuation Tools That Unlock Business Wealth*; John Wiley & Sons, Inc.: Hoboken, NJ, USA, 2010. Available online: https://books.google.co.uk/books?id=qfADF9jgylcC&pg=PA133&lpg=PA133&dq=value+drivers+scholar&source=bl&ots=GqG_Gzu-YG&sig=Yz98aw-bcYoZerffhRSQPKpvw4E&hl=en&sa=X&ved=0ahUKEwjKhfGxz-TOAhUsIMAKHSONBA8Q6AEIITAB#v=onepage&q=value%20drivers%20scholar&f=false (accessed on 27 July 2016).
37. Ohlson, J.A. Earnings, Book Values, and Dividends in Equity Valuation. *Contemp. Account. Res.* **1995**, *11*, 661–687, doi:10.1111/j.1911-3846.1995.tb00461.x.
38. Ohlson, J.A. Earnings, book values, and dividends in equity valuation: An empirical perspective. *Contemp. Account. Res.* **2001**, *18*, 107–120, doi:10.1506/7TPJ-RXQN-TQC7-FFAE.
39. Ohlson, J.A. On Accounting-Based Valuation Formulae. *Rev. Account. Stud.* **2005**, *10*, 323–347, doi:10.1007/s11142-005-1534-4.
40. Penman, S. *Financial Statement Analysis and Security Valuation*, 4th ed.; McGraw-Hill: London, UK, 2010.
41. Penman, S.H. On Comparing Cash Flow and Accrual Accounting Models for Use in Equity Valuation: A Response to Lundholm and O'Keefe (CAR, Summer 2001). *Contemp. Account. Res.* **2001**, *18*, 681–692, doi:10.1506/DT0R-JNEG-QL60-7CBP.
42. Penman, S.H.; Sougiannis, T. A Comparison of Dividend, Cash Flow, and Earnings Approaches to Equity Valuation. *Contemp. Account. Res.* **1998**, *15*, 343–383, doi:10.1111/j.1911-3846.1998.tb00564.x.
43. Richardson, G.; Tinaikar, S. Accounting Based Valuation Models: What Have We Learned? *Account. Financ.* **2004**, *44*, 223–255, doi:10.1111/j.1467-629X.2004.00109.x.
44. Sainsbury's. Sainsbury's about Us. 2016. Available online: https://www.about.sainsburys.co.uk/about-us/our-purpose (accessed on 30 June 2016).
45. Shaked, I.; Michel, A.; Leroy, P. Creating Value Through EVA—Myth or Reality? *Strategy Bus.* **1997**, *4*. Available online: https://www.strategy-business.com/article/12756?gko=3ba97 (accessed on 22 July 2016).
46. Shi, J.; Ausloos, M.; Zhu, T. Benford's law first significant digit and distribution distances for testing the reliability of financial reports in developing countries. *Phys. A* **2018**, *492*, 877–888, doi:10.1016/j.physa.2017.11.017.
47. Sofat, R.; Hiro, P. *Strategic Financial Management*, 2nd ed.; PHI Learning Private Limited: New Delhi, India, 2016. Available online: https://books.google.co.uk/books?id=AtDKCgAAQBAJ&printsec=frontcover&dq=Strategic+Financial+Management&hl=en&sa=X&ved=0ahUKEwiO2sf-1uTOAhVYFMAKHUEyBs4Q6AEIMjAB#v=onepage&q=Strategic%20Financial%20Management&f=false (accessed on 2 August 2016).
48. Stern, J.; Stewart, G.; Chew, D. The EVA Financial Management System. *J. Appl. Corp. Financ.* **1995**, *8*, 32–46, doi:10.1111/j.1745-6622.1995.tb00285.x.
49. Tesco. Tesco about Us Our Business. 2016. Available online: https://www.tescoplc.com/about-us/ (accessed on 12 June 2016).
50. Yahoo Finance. Marks & Spencer Share Price History. 2016. Available online: https://uk.finance.yahoo.com/q/hp?s=MKS.L&b=31&a=02&c=2016&e=31&d=02&f=2016&g=d (accessed on 10 August 2016).
51. Young, S.D.; O'Byrne, S.F. *EVA and Value-Based Management: A Practical Guide to Implementation*; McGraw-Hill Education: New York, NY, USA, 2000.
52. Yi, R.; Chang, Y.W.; Xing, W.; Chen, J. Comparing relative valuation efficiency between two stock markets. *Q. Rev. Econ. Financ.* **2019**, *72*, 159–167, doi:10.1016/j.qref.2018.11.008.